\documentclass[superscriptaddress,groupedaddress,nofootnoteinbib,12pt]{article} 
\usepackage{graphicx}
\usepackage{color}
\usepackage{dcolumn}   
\usepackage{bm}     
\usepackage{bbm}       
\usepackage{amssymb}  
\usepackage{amsmath}
\usepackage{sectsty}
\usepackage{latexsym}
\usepackage{float}
\usepackage{ifthen}
\usepackage{caption,subfig}
\usepackage{enumerate}
\usepackage{url}
\usepackage{caption,subfig}
\usepackage{jcappub}
\usepackage{amsopn}

\usepackage{amsfonts}
\usepackage{multirow}
\usepackage{array}
\usepackage{booktabs}
\usepackage{rotating}
\usepackage{tipa}

\def\clap#1{\hbox to 0pt{\hss#1\hss}}

\def\({\left(}
\def\){\right)}
\def\[{\left[}
\def\]{\right]}
\def\bea{\begin{eqnarray}}

\def\eea{\end{eqnarray}}
\def\be{\begin{equation}}
\def\ee{\end{equation}}
\def\mpl{M_{\rm Pl}}

\newcommand{\Diff}{{\rm D}}
\newcommand{\diff}{{\rm d}}
\newcommand{\TT}{\mathbb{T}}
\newcommand{\TTGR}{\mathring{\mathbb{T}}}

\newcommand{\QQ}{\mathbb{Q}}
\newcommand{\QQGR}{\mathring{\mathbb{Q}}}

\newcommand{\Lag}{\mathcal{L}}

\newcommand{\mR}{\mathcal{R}}

\def\clap#1{\hbox to 0pt{\hss#1\hss}}

\newcommand{\e}{\mathrm{e}}
\newcommand{\ie}{\text{\textschwa}}

\def\TT{\mathbb{T}}
\def\QQ{\mathbb{Q}}

\newcommand{\mS}{\mathcal{S}}

\definecolor{forestgreen}{rgb}{0.133,0.545,0.133}
\newboolean{editorial}
\setboolean{editorial}{true}
\newcommand{\editorial}[2]{\ifthenelse{\boolean{editorial}}{\textcolor{red}{[\textsf{\textbf{{#1}}}: }\textcolor{blue}{\textsf{{#2}}}\textcolor{red}{]}}{}}

\renewcommand{\vec}[1]{\bm{\mathrm{{#1}}}}

 \def\be   {\begin{equation}}   \def\ee   {\end{equation}}

 \def\ba  {\begin{eqnarray}}   \def\ea  {\end{eqnarray}}

\renewcommand{\thesubsection}{\arabic{section}.\arabic{subsection}}

\begin{document}

\title{The Geometrical Trinity of Gravity}

\author{Jose Beltr\'an Jim\'enez$^{a}$, Lavinia Heisenberg$^b$ and Tomi S. Koivisto$^{c,d}$}
\affiliation{$^a$Departamento de F\'isica Fundamental, Universidad de Salamanca, E-37008 Salamanca, Spain.\\
$^b$Institute for Theoretical Physics, 
ETH Zurich, Wolfgang-Pauli-Strasse 27, 8093, Zurich, Switzerland\\
$^c$ Nordita, KTH Royal Institute of Technology and Stockholm University, Roslagstullsbacken 23, SE-10691 Stockholm, Sweden \\
$^d$ National Institute of Chemical Physics and Biophysics, R\"avala pst. 10, 10143 Tallinn, Estonia}

	\emailAdd{jose.beltran@usal.es}
	\emailAdd{lavinia.heisenberg@phys.ethz.ch}
	\emailAdd{tomik@astro.uio.no}

\abstract{The geometrical nature of gravity emerges from the universality dictated by the equivalence principle. In the usual formulation of General Relativity, the geometrisation of the gravitational interaction is performed in terms of the spacetime curvature, which is now the standard interpretation of gravity. However, this is not the only possibility. In these notes we discuss two alternative, though equivalent, formulations of General Relativity in flat spacetimes, in which gravity is fully ascribed either to torsion or to non-metricity, thus putting forward the existence of three seemingly unrelated representations of the same underlying theory. Based on these three alternative formulations of General Relativity, we then discuss some extensions.}

\maketitle

\section{Introduction}
Gravity and geometry have accompanied each other from the very conception of General Relativity (GR) brilliantly formulated by Einstein in terms of the spacetime curvature. This inception of identifying gravity with the curvature has since grown so efficiently that it is now a common practice to recognise the gravitational phenomena as a manifestation of having a curved spacetime. As Einstein ingeniously envisioned, the existence of a geometrical formulation of gravity is granted by the equivalence principle that renders the gravitational interaction oblivious to the specific type of matter and hints towards an intriguing relation of gravity with inertia. Thus, the motion of particles can be naturally associated to the geometrical properties of spacetime. If we embrace the geometrical character of gravity advocated by the equivalence principle, it is pertinent to explore in which equivalent manners gravity can be geometrised. It is then convenient to recall at this point that a spacetime can be endowed with a metric and an affine structure \cite{Schroedinger1950} determined by a metric tensor $g_{\mu\nu}$ and a connection $\Gamma^\alpha{}_{\mu\nu}$, respectively. These two structures, although completely independent, enable the definition of geometrical objects that allow to conveniently classify geometries. The failure of the connection to be metric is encoded in the non-metricity
\be
Q_{\alpha\mu\nu}\equiv \nabla_\alpha g_{\mu\nu},
\ee
while its antisymmetric part defines the torsion
\be
T^\alpha{}_{\mu\nu}\equiv 2\Gamma^\alpha{}_{[\mu\nu]}.
\ee
Among all possible connections that can be defined on a spacetime, the Levi-Civita connection is the unique connection that is symmetric and metric-compatible. These two conditions fix the Levi-Civita connection to be given by the Christoffel symbols of the metric
\be \label{levi-civita}
\left\{^{\phantom{i} \alpha}_{\mu\nu}\right\}=\frac12 g^{\alpha\lambda}\big(g_{\lambda\nu,\mu}+g_{\mu\lambda,\nu}-g_{\mu\nu,\lambda}\big).
\ee
The corresponding covariant derivative will be denoted by $\mathcal{D}$ so that we will have $\mathcal{D}_\alpha g_{\mu\nu}=0$.  A general connection $\Gamma^\alpha{}_{\mu\nu}$ then admits the following convenient decomposition:
\be
\Gamma^\alpha_{\phantom{\alpha}\mu\nu}=\left\{^{\phantom{i} \alpha}_{\mu\nu}\right\} + K^\alpha_{\phantom{\alpha}\mu\nu} + L^\alpha_{\phantom{\alpha}\mu\nu}
\label{Eq:decomposition}
\ee
with 
\ba
K^\alpha_{\phantom{\alpha}\mu\nu}= \frac{1}{2}T^\alpha_{\phantom{\alpha}\mu\nu} + T_{(\mu{\phantom{\alpha}\nu)}}^{\phantom{,\mu}\alpha}\,,\quad
L^\alpha_{\phantom{\alpha}\mu\nu}= \frac{1}{2} Q^{\alpha}_{\phantom{\alpha}\mu\nu} - Q_{(\mu\phantom{\alpha}\nu)}^{\phantom{(\mu}\alpha}
\ea
the contortion and the disformation pieces, respectively. Notice that, while the Levi-Civita part is non-tensorial, the contortion and the disformation have tensorial transformation properties under changes of coordinates.

The curvature is determined by the usual Riemann tensor 
\be \label{curvature}
R^\alpha{}_{\beta\mu\nu}(\Gamma)=\partial_\mu \Gamma^\alpha{}_{\nu\beta}-\partial_\nu \Gamma^\alpha{}_{\mu\beta}+\Gamma^\alpha{}_{\mu\lambda}\Gamma^\lambda{}_{\nu\beta}-\Gamma^\alpha{}_{\nu\lambda}\Gamma^\lambda{}_{\mu\beta}.
\ee
A relation that will be useful later on is how the Riemann tensor transforms under a shift of the connection $\hat{\Gamma}^\alpha_{\phantom{\alpha}\mu\nu} = {\Gamma}^\alpha_{\phantom{\alpha}\mu\nu} + \Omega^\alpha_{\phantom{\alpha}\mu\nu}$, with $\Omega^\alpha{}_{\mu\nu}$ an  arbitrary tensor. Under such a shift, the Riemann tensor becomes
\be 
 \hat{R}^\alpha_{\phantom{\alpha}\beta\mu\nu}=R^\alpha_{\phantom{\alpha}\beta\mu\nu} + T^\lambda_{\phantom{\lambda}\mu\nu}\Omega^\alpha_{\phantom{\alpha}\lambda\beta} + 2\nabla_{[\mu}\Omega^\alpha_{\phantom{\alpha}\nu]\beta} +2\Omega^\alpha_{\phantom{\alpha}[\mu\lvert\lambda\rvert}\Omega^\lambda_{\phantom{\lambda}\nu]\beta}\,.
 \label{eq:RtoR}
\ee
where $ \hat{R}^\alpha_{\phantom{\alpha}\beta\mu\nu}$ and $R^\alpha_{\phantom{\alpha}\beta\mu\nu}$ are the Riemann tensors of $\hat{\Gamma}$ and $\Gamma$ respectively and $\nabla$ is the covariant derivative associated to $\Gamma$.

After gathering the relevant geometrical objects, we can use them to characterise a spacetime as follows:
\begin{itemize}
\item Metric: the connection is metric-compatible, $Q_{\alpha\mu\nu}(\Gamma,g)=0$. Non-metricity measures how much the length of vectors change as we parallel transport them, so in metric spaces the length of vectors is conserved.
\item Torsionless: the connection is symmetric, $T^\alpha{}_{\mu\nu}(\Gamma)=0$. Torsion gives a measure of the non-closure of the parallelogram formed when two infinitesimal vectors are parallel transported along each other. For this reason it is usually said that parallelograms do not close in the presence of torsion.
\item Flat: the connection is not curved, $R^\alpha{}_{\beta\mu\nu}(\Gamma)=0$. Curvature measures the rotation experienced by a vector when it is parallel transported along a closed curve. This represents an obstacle to compare vectors defined at different spacetime points. In flat spaces however vectors do not rotate as they are transported so that there is a better notion of parallelism at a distance. This is the reason why theories formulated in these spaces are referred to as teleparallel.
\end{itemize}

\vspace{-1cm}
\begin{center}
\begin{figure}[h]
\includegraphics[width=17cm]{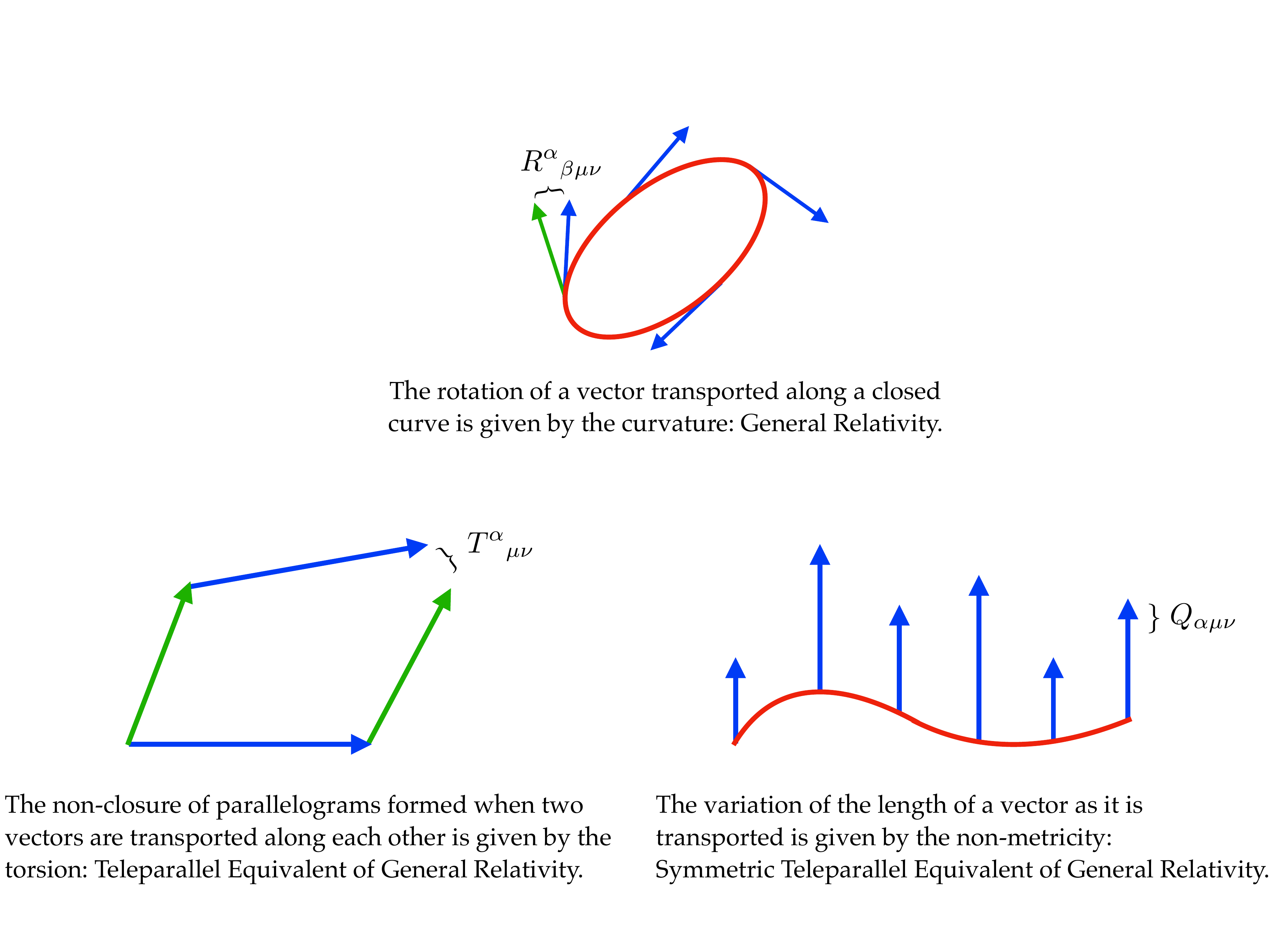}
   \vspace{-1.25cm}
  \caption{This figure illustrates the geometrical meaning of the curvature, the torsion and the non-metricity when the remaining objects vanish. We should emphasise that when a vector is transported along a closed curve in a general geometry, it will acquire a rotation determined by $R_{[\alpha\beta]\mu\nu}$ and a length variation given by $R_{(\alpha\beta)\mu\nu}$. It should be compared to Fig. \ref{FigA} where it is summarised how General Relativity admits equivalent representations in terms of these three geometrical objects. }
 \label{geometricalObjects}
\end{figure}
\end{center}

Einstein's original formulation founded GR on a metric and torsionless spacetime and imputed gravity to the curvature. It is however natural to explore, as did also Einstein later, if gravity can instead be ascribed to the remaining attributes that a spacetime can have, i.e., to the torsion and to the non-metricity. In these notes, we will confirm that the very same underlying theory, i.e., GR, can be equivalently described in terms of these three seemingly unrelated elements, knocking into shape a geometrical trinity of gravity. We will nonetheless illustrate some subtle, conceptual and practical, differences among them.\\

\section{General Relativity}
Before delving into its alternative representations, let us start with the best-known formulation of GR where gravity is identified with the curvature of spacetime and the dynamics is described by the Hilbert action 
\be
\mathcal{S}_{\rm GR_{(2)}}=\frac{1}{16\pi G}\int\diff^4x\sqrt{-g}\,\mathcal{R}(g)
\label{eq:EHaction}
\ee
with $\mathcal{R}=g^{\mu\nu} R_{\mu\nu}(\{\})$ the curvature of the Levi-Civita connection \eqref{levi-civita}. The fundamental object here is the metric with its ten components, but the four-parameter Diffs (diffeomorphisms) gauge symmetry reduces them to only two dynamical degrees of freedom (dof), as it corresponds to a massless spin 2 particle. At a more technical level, the constraints on the connection being symmetric and torsion-free should more properly be incorporated by adding suitable Lagrange multiplier fields enforcing such constraints 
\be
\mathcal{S}_{\rm GR_{(1)}}=\int\diff^4x\left[ \frac{\sqrt{-g}}{16\pi G}g^{\mu\nu}R_{\mu\nu}(\Gamma) + \lambda_{\alpha}{}^{\mu\nu}T^\alpha{}_{\mu\nu} + \hat{\lambda}^{\alpha}{}_{\mu\nu}Q_{\alpha}{}^{\mu\nu}\right]\,,
\label{eq:EHaction1}
\ee
where the $R$ is now the scalar of the curvature (\ref{curvature}). Being the two constraints imposed on the connection integrable and holonomic, we can simply solve them, insert back into the action (\ref{eq:EHaction1}) and obtain the second order action for the metric (\ref{eq:EHaction}). There is however a very remarkable property of the Hilbert action that makes it special. When considered in the metric-affine formalism, i.e., with a completely general connection not fixed a priori nor by means of Lagrange multipliers, the first term in the action (\ref{eq:EHaction1}) alone gives rise to equations for the connection that fix it to be precisely the Levi-Civita connection of the spacetime metric $g_{\mu\nu}$. A subtlety of this result is related to the existence of a projective symmetry for the Hilbert action. In fact, under a projective transformation of the connection $\delta_\zeta\Gamma^\alpha_{\mu\beta}=\zeta_\mu\delta^\alpha{}_\beta$, the Riemann tensor changes as $\delta_\zeta R^\alpha{}_{\beta\mu\nu}=2\delta^\alpha{}_\beta\partial_{[\mu}\zeta_{\nu]}$ so that the Ricci scalar $R$ is invariant. As a consequence, the projective mode is left undetermined by the field equations as a gauge mode which can then be fixed by simply making a projective gauge choice (see e.g. \cite{reviewBI} and references therein for more details). 

This formulation has some inherent difficulties owed to working in a curved spacetime. Among others, the Hilbert action (\ref{eq:EHaction}) contains second derivatives of the metric so the variational principle is not well-posed in the usual sense since one is led to fix normal derivatives of the metric on the boundary, which further hinders a composition law for the path integral. As it is well-known, these formal issues are solved by the Gibbons-Hawking-York (GHY) boundary term \cite{Gibbons:1976ue},  whose physical importance is prominently reflected by the fact that it entirely determines the black hole entropy. 

\section{Metric teleparallelism}
\label{tele}

An alternative geometrical framework, attributing gravity to the torsion, is defined by its flatness and metric compatibility. 
These properties conform to the {Weitzenb\"ock} connection characteristics. As the natural starting point to construct the theories, we will consider the most general even-parity second order quadratic form that can be built with the torsion and which is given by the three-parameter combination
\be
\TT\equiv-\frac{c_1}{4} T_{\alpha\mu\nu}T^{\alpha\mu\nu}-\frac{c_2}{2}T_{\alpha\mu\nu}T^{\mu\alpha\nu}+c_3T_\alpha T^\alpha\,,
\label{Eq:defTT}
\ee
where $c_1,c_2,c_3$ are some free parameters and $T_\mu=T^\alpha{}_{\mu\alpha}$ is the trace of the torsion. At the level of the action, the constraints will be enforced by introducing suitable Lagrange multipliers so that the general quadratic action is given by
\be \label{abcd}
\mS_{\TT}   =- \int\diff^4x\left[\frac{1}{16\pi G}\sqrt{-g}\,\mathbb{T}  + 
\lambda_\alpha^{\phantom{\alpha}\beta\mu\nu} R^\alpha_{\phantom{\alpha}\beta\mu\nu} + \hat{\lambda}^\alpha_{\phantom{\alpha}\mu\nu}\nabla_\alpha g^{\mu\nu} \right]\,.
\ee
Notice that the Lagrange multipliers have the obvious symmetries $\lambda_{\alpha}^{\phantom{\alpha}\mu\beta\nu}  = \lambda_{\alpha}^{\phantom{\alpha}\mu[\beta\nu]}$, $\hat{\lambda}^\alpha_{\phantom{\alpha}\mu\nu}=\hat{\lambda}^\alpha_{\phantom{\alpha}(\mu\nu)}$, and that we have defined them as tensorial densities of weight $-1$ for convenience.

We will start by solving the constraints. Since the curvature is the field strength of the connection, its vanishing implies that the connection must be a pure gauge field or, in other words, the connection is purely inertial. It can then be parameterised by an element $\Lambda^\alpha{}_\mu$ of the general linear group $GL(4,\mathbb{R})$ as 
\be \label{pure}
\Gamma^\alpha{}_{\mu\nu}=(\Lambda^{-1})^\alpha{}_\lambda\partial_\mu \Lambda^\lambda{}_\nu\,.
\ee
The metric constraint further restricts $\Lambda^\alpha{}_\beta$ and the metric to satisfy the following relation\footnote{Let us note here the typo in the corresponding Eq.(50) in Ref. \cite{BeltranJimenez:2018vdo} where the inverse was misplaced.}:
\be
2(\Lambda^{-1})^\lambda{}_\kappa\partial_\alpha\Lambda^\kappa{}_{(\mu} g_{\nu)\lambda} =\partial_\alpha g_{\mu\nu}\,. 
\ee
This equation determines the metric in terms of the connection, i.e., given a solution for $\Lambda^\alpha{}_\mu$, the above relation will determine the corresponding metric. This is analogous to a vierbein determining the metric, as will be shortly clarified in Sec. \ref{vierbein}. These are all the required elements to formulate gravity in terms of the torsion, given thus by 
\be
T^\alpha{}_{\mu\nu}=2(\Lambda^{-1})^\alpha{}_\lambda\partial_{[\mu} \Lambda^\lambda{}_{\nu]}\,.
\label{eq:TorsionLambda}
\ee

We now proceed to explore if GR can be recovered by a suitable choice of parameters. This is in fact possible in a simple manner by noticing that, when non-metricity vanishes, we can use (\ref{eq:RtoR}) to express the Ricci scalar as
\be
R = \mathcal{R}(g) + \mathring{\mathbb{T}} + 2\mathcal{D}_\alpha T^\alpha\,, 
\ee
where $\TTGR$ is nothing but $\TT$ setting $c_1=c_2=c_3=1$ in Eq.(\ref{Eq:defTT}). The flatness condition $R=0$ then tells us that the Ricci scalar $\mathcal{R}$ of the Levi-Civita connection differs from $\TTGR$ by a total derivative. We can thus conclude that the dynamics of GR is identically recovered by 
\be \label{ab}
\mathcal{S}_{\rm TEGR}=-\frac{1}{16\pi G}\int\diff^4x\sqrt{-g}\,\TTGR(g,\Lambda)\,.
\ee
The resulting theory is the well-known Teleparallel Equivalent of GR (TEGR). The usual formulation of TEGR makes fundamental use of the tetrad fields, which requires additional geometrical structure to introduce the frame bundle and the corresponding soldering form \cite{Aldrovandi:2013wha}. Here, instead, the same theory results from a manifestly covariant approach. 

\subsection{Vierbein formulation}
\label{vierbein}

For comparison, we can briefly review the vierbein formulation of GR and TEGR. 
The vierbein is introduced as a set of covectors $\ie_a=\ie_a{}^\mu\partial_\mu$ that are orthonormal with respect to the Minkowski metric in the sense that $\eta_{ab}\e^a{}_{\mu} \e^b{}_{\nu}=g_{\mu\nu}$, where the set of 1-forms $\e^a= \e^a{}_\mu\diff x^\mu$ is the inverse vierbein. These objects naturally live in the Lorentz frame bundle which is endowed with the usual spin connection $\omega^a{}_b=\omega^a{}_{\mu b}\diff x^\mu$. The vierbein is related to the translation gauge potential $B^a=B^a{}_\mu\diff x^\mu$ as $\e^a = B^a + \Diff \xi^a$, where $\Diff$ is the covariant exterior derivative with respect to the spin connection. It is necessary to introduce the field $\xi^a$, which can be interpreted as the tangent space coordinate  \cite{Aldrovandi:2013wha}, since the vierbein has the covariant transformation law, but $B^a$ transforms as a connection. The field strength of translations, $\Diff B^a$, coincides with the torsion two-form, $T^a = \Diff\e^a = \diff \e^a+\omega^a{}_b \wedge \e^b$, if there is no curvature.
The curvature is, as usual, $R^a{}_b=\Diff \omega^a{}_b = \diff \omega^a{}_b+\omega^a{}_c\wedge \omega^c{}_b$, and it can be understood as the field strength of the Lorentz rotations. 
An important relation is that between the affine connection and the spin connection, $\Gamma^\alpha{}_{\mu\nu} = \ie_a{}^\alpha \Diff_\mu \e^a{}_\nu =  -\e^a{}_\nu \Diff_\mu \ie_a{}^\alpha$. Taking this into account, the spacetime tensors are related to the field strength two-forms simply as $R^\alpha{}_{\beta\mu\nu}=\ie_a{}^\alpha
R^a{}_{b\mu\nu}\e^b{}_\beta$ and $T^\alpha{}_{\mu\nu}=\ie_a{}^\alpha T^a{}_{\mu\nu}$.

We have now all the ingredients to rewrite our actions for GR and TEGR in the vierbein formalism. To obtain the second order vierbein formulation of the Hilbert action, we would simply
insert the definition of the metric into (\ref{eq:EHaction}), and write $\mathcal{R}(\ie_a)$ instead of $\mathcal{R}(g)$, and employ the determinant $\ie$ which is the same as $\sqrt{-g}$. 
The proper first order formulation of GR would be obtained instead by doing the corresponding replacements in (\ref{eq:EHaction1}) (where the last constraint is unnecessary in the Lorentz bundle,
but would be needed in the general linear bundle). Solving the spin connection from the constraint of vanishing torsion, we would obtain the non-trivial expression for the $\omega^a{}_b$ that boils down to the Eq.(\ref{levi-civita}), when rewritten for the spacetime affine connection. 

In the teleparallel formulation, in contrast, there exists a solution with $\omega^a{}_b=0$. Again, we may begin with the action (\ref{abcd}) with the respective replacements such that the torsion is understood as a function $\mathbb{T}(\ie,\omega)$. 
The flatness condition can be solved and it determines the spin connection to be $\omega^a{}_{ b}=(\Lambda^{-1})^a{}_c\diff \Lambda^c{}_b$, in analogy
with (\ref{pure}).
When plugged back into the action and choose the TEGR parameter combination $c_1=c_2=c_3=1$, we obtain (\ref{ab}), wherein now $\TTGR=\TTGR (\ie_a,\Lambda^a{}_b)$. This formulation was introduced in \cite{Golovnev:2017dox} and the physical interpretation of the purely inertial spin connection determined by the matrix $\Lambda^a{}_b$ was clarified in the recent review \cite{Krssak:2018ywd}. Here instead, we would like to only make the further remark on the relation to the formulation in terms of the affine connection. If we choose the solution $\omega^a{}_b=0$,
which could be called the Weitzenb\"ock gauge,  the torsion is then determined by the vierbein as $T^a{}_{\mu\nu}=\partial_{[\mu}\e^a{}_{\nu]}$ and we can project it with the vierbein to obtain $T^\alpha{}_{\mu\nu}=\ie^\alpha{}_a\partial_{[\mu}\e^a{}_{\nu]}$. Comparing with our (\ref{eq:TorsionLambda}) in the metric-affine formalism, it is interesting to see that the gauge transformation of the vanishing affine connection had essentially generated the vierbein in the Weitzenb\"ock gauge.

One can identify the TEGR as a special case among the family of quadratic theories described by $\TT$ because it features an additional local Lorentz symmetry: we may transform only the
vierbein and neglect the spin connection. 
 This local symmetry is only realised up to a total derivative, what has some important consequences which were reviewed in Ref. \cite{Krssak:2018ywd}. Consequently, out of the 16 components of the vierbein, or of the $\Lambda^\alpha{}_\mu$ in the covariant formulation, 8 are non-dynamical due to Diffs, while 6 more simply reflect the freedom in performing a Lorentz transformation, leaving thus the 2 dynamical dof's of GR.

\subsection{Alternative theories}

The metric teleparallel reformulation of GR can be straightforwardly extended in two different directions, both of which result in the loss of symmetries. The first modification consists in leaving the three parameters in (\ref{Eq:defTT}) free, which is known as New GR. In that case, the extra local Lorentz symmetry disappears and this results in the appearance of additional propagating fields. 
As a first check of the content of this extension, we can look at the linearised theory around Minkowski. This has been performed in the formulation that makes use of the vierbeins. It would be interesting to redo the analysis in the covariant formalism presented here without resorting to the vierbein formalism. Of course, we expect to obtain the same field content. The perturbed vierbein around Minkowski is simply $\e^a{}_\mu=\delta^a{}_\mu+ A^a{}_\mu$. The background configuration with $\delta^a{}_\mu$ allows to construct\footnote{This simply means that we can identify the tangent space and the curved indices at first order.} $A_{\mu\nu}\equiv \delta^a{}_\mu A^a{}_\nu$. This perturbation can then be decomposed into its symmetric $h_{\mu\nu}=2A_{(\mu\nu)}$ and antisymmetric $b_{\mu\nu}\equiv 2A_{[\mu\nu]}$ pieces. The quadratic action for these fields was given in Eq.(4.173) of Ref. \cite{Ortin:2015hya}, and a more general case was studied in
Ref. \cite{Koivisto:2018loq}. An essential piece in the action is the coupling of the two pieces of perturbations,
\be
L_{h-b} = -\frac{1}{16}\left( c_1 + c_2 - 2c_3\right) h^{\mu\nu}\partial_\mu \partial^\alpha b_{\alpha\nu}\,, 
\ee
which vanishes if $c_1+c_2=2c_3$. The consistency of the theory requires this. Up to the overall normalisation, imposing this condition leaves a one-parameter class of theories which propagates, in addition to the graviton, a Kalb-Ramond field. The latter is removed if one further imposes that $c_1+2c_3=3c_2$, which leaves the special case of TEGR \cite{Ortin:2015hya}.
It is interesting that the crucial constraint $c_1+c_2=2c_3$ is related to a symmetry which renders the inverse vierbein equivalent to the translation gauge potential \cite{Koivisto:2018loq}.
In the following Section we will uncover another perspective, from symmetric teleparallelism, to the relevance of making the theory oblivious to the $\xi^a$. 

Another straightforward modification is simply taking non-linear extensions of the TEGR action, which results in the so-called $f(\TTGR)$ theories. Since the local Lorentz symmetry is realised up to a total derivative, these extensions also lose such a symmetry and additional dofs are expected. On details of these theories, we refer the reader to \cite{Golovnev:2017dox,Krssak:2018ywd} and their references.

\section{Symmetric Teleparallelisms}
The advent of GR fully ascribed to the non-metricity is materialised in a flat and torsion free geometry \cite{Nester:1998mp}. As we will see, the geometrical framework for this formulation of GR is arguably the simplest among the three equivalent representations because there is no curvature nor torsion and non-metricity is left as the fundamental geometrical object. Furthermore, the connection can be globally\footnote{Barring possible topological obstructions.} completely removed by an appropriate choice of coordinates so that the spacetime is trivially connected. 

\subsection{Symmetric Teleparallel Equivalent of GR: Coincident GR}
Once we have described the coincident GR's dwell, we will proceed as before considering the most general even-parity second order quadratic form of the non-metricity \cite{Einstein}
\be
\QQ =   \frac{c_1}{4}Q_{\alpha\beta\gamma}Q^{\alpha\beta\gamma} -  \frac{c_2}{2}Q_{\alpha\beta\gamma}Q^{\beta\alpha\gamma} 
  -   \frac{c_3}{4}Q_\alpha Q^\alpha  +(c_4-1)\tilde{Q}_\alpha \tilde{Q}^\alpha
  + \frac{c_5}{2}Q_\alpha\tilde{Q}^\alpha\,,
\ee
where $Q_\alpha=Q_{\alpha\lambda}{}^\lambda$ and $\tilde{Q}_\alpha=Q^\lambda{}_{\lambda\alpha}$ are the two independent traces of the non-metricity. The general quadratic action including suitable Lagrange multipliers is then
\be \label{abcd2}
\mS_{\QQ}=-\int\diff^4x\left[\frac{1}{16\pi G}\sqrt{-g}\,\QQ+\lambda_{\alpha}{}^{\beta\mu\nu} R^\alpha{}_{\beta\mu\nu}+\lambda_\alpha{}^{\mu\nu} T^\alpha{}_{\mu\nu}\right]\,.
\ee
In this case we have a 5-parameter family of quadratic theories. We could now explore the whole space of theories and check the existence of some particular case that gives rise to an equivalent of GR. However, we can again show the existence of an equivalent to GR by using (\ref{eq:RtoR}) for a torsion-free connection, which gives
\be
R   =  \mathcal{R}(g)  + \QQGR +   \mathcal{D}_\alpha ( Q^\alpha - \tilde{Q}^\alpha )\,,
\label{eq:RtoQ}
\ee
where $\QQGR$ is given by $\QQ$ setting all $c_i=1$ so that, in a flat spacetime with $R=0$, we find the relation $\mathcal{R}(g)=-\QQGR-\mathcal{D}_\alpha ( Q^\alpha - \tilde{Q}^\alpha )$ and, consequently, the action 
\be
\mathcal{S}_{\rm STGR}=-\frac{1}{16\pi G}\int\diff^4x\sqrt{-g}\,\QQGR(g,\xi)\,,
\label{Eq:CGRaction}
\ee
where STGR stands for Symmetric Teleparallel Equivalent of GR, differs from the Hilbert action by a total derivative, thus reproducing the dynamics of GR. As in the TEGR, the quadratic form $\QQGR$ is special because it has an enhanced symmetry that is realised up to a total derivative. This will become clearer in a moment, but now let us look at the affine structure of this theory in more detail by solving the constraints. The flatness condition again restricts the connection to be purely inertial so that it can be parameterised by a general element $\Lambda^\alpha{}_\beta$ of $GL(4\mathbb{R})$. This form of the connection in combination with the absence of torsion leads to the additional constraint $\partial_{[\mu}\Lambda^\alpha{}_{\nu]}=0$. The general element of $GL(4,\mathbb{R})$ determining the connection can thus be parameterised by a set of functions $\xi^\lambda$ so that\footnote{Of course, in this expression $\frac{\partial x^\alpha}{\partial\xi^\lambda}$ should be interpreted as the inverse of the matrix $\frac{\partial \xi^\lambda}{\partial x^\alpha}$.}
\be
\Gamma^\alpha{}_{\mu\nu}=\frac{\partial x^\alpha}{\partial\xi^\lambda}\partial_\mu\partial_\nu\xi^\lambda\,.
\ee
This seemingly innocent form of the connection hides however an outstanding property of the non-metricity representation of GR, namely: the connection can be trivialised by a coordinate transformation \cite{BeltranJimenez:2017tkd}. 
The gauge where the connection vanishes gives $\xi^\alpha=x^\alpha$, which can be interpreted as the gauge where the origin of the tangent space parameterised by $\xi^\alpha$ coincides with the spacetime origin and for this property it is dubbed the {\it coincident gauge}\footnote{This gauge is defined up to an affine transformation $x^\mu\rightarrow ax^\mu+b$ with $a$ and $b$ constants. Since this residual global symmetry does not vanish at infinity, it might lead to interesting properties of the infrarred structure of the theory.}. 

An interesting form of writing the STEGR action is in terms of the disformation directly as
\be \label{el}
\mS_{\rm STEGR}=\frac{1}{16\pi G}\int\diff^4x\sqrt{-g}g^{\mu\nu}\Big( L^\alpha{}_{\beta\mu}L^\beta{}_{\nu\alpha} - L^\alpha{}_{\beta\alpha}L^\beta{}_{\mu\nu}\Big)\,.
\ee
The interest of this expression is that, in the coincident gauge and recalling the decomposition (\ref{Eq:decomposition}), the triviality of the connection directly gives the relation 
\be
\left\{^{\phantom{i} \alpha}_{\mu\nu}\right\}=-L^\alpha_{\phantom{\alpha}\mu\nu}.
\ee
It is then straightforward to verify that, in this coincident gauge, the action (\ref{Eq:CGRaction}) can be written as
\be
\mathcal{S}_{\rm CGR}=\mS_{\rm STEGR}[\Gamma=0]=\frac{1}{16\pi G}\int\diff^4x\sqrt{-g}g^{\mu\nu}\Big(\left\{^{\phantom{i} \alpha}_{\beta\mu}\right\} \left\{^{\phantom{i} \beta}_{\nu\alpha}\right\} -\left\{^{\phantom{i} \alpha}_{\beta\alpha}\right\}\left\{^{\phantom{i} \beta}_{\mu\nu}\right\} \Big)\,.
\ee
We call this the action of {\it Coincident GR}. Remarkably, it reproduces the Einstein action for GR consisting of the Hilbert action devoid of boundary terms. It has the advantage of only involving first derivatives of the metric, thus leading to a well-posed variational principle without any GHY boundary terms. However, Diff invariance is only realised up to a total derivative which causes the action to depend on the chosen coordinates. It may look striking that we refer to Diffs even though we have used them to fix the coincident gauge, but there is no onus. The reason is that, similarly to the TEGR being special because it features an additional symmetry, the theory (\ref{Eq:CGRaction}) is special among the quadratic theories because it enjoys an enhanced four-parameter gauge symmetry so the full theory actually has an eight-parameter gauge symmetry. In the coincident gauge, the additional symmetry appears as a Diff symmetry. Furthermore, unlike the TEGR where the metric and the connection are related, in the non-metricity formulation of GR, the connection is fundamentally pure gauge and all the dynamics can be encoded into the metric, now in a trivially connected spacetime. In this respect, it is worth to point out that the fields $\xi^\alpha$ that parameterise the connection play in turn the role of St\"uckelberg fields associated to coordinates transformations invariance and the coincident gauge is nothing but the corresponding unitary gauge.

\subsection{General quadratic theory}
As for the TEGR, there are two straightforward extensions that can be considered. The first one corresponds to considering arbitrary parameters in the general quadratic action, in which case the Diffs in the coincident gauge is lost, thus resulting in additional dofs. This family of theories was dubbed Newer General Relativity. It is illustrative to look at the structure of these theories around a Minkowski background with $g_{\mu\nu}=\eta_{\mu\nu}+h_{\mu\nu}$. Unlike in the TEGR, there is no antisymmetric field and the whole dynamics is encapsulated into $h_{\mu\nu}$. The quadratic action then reads
\be \label{h_action}
\mathcal{L}=\frac{c_1}{4}\partial_\alpha h_{\mu\nu} \partial^\alpha h^{\mu\nu}-\left(\frac{c_2}{2}+1-c_4\right)\partial_\alpha h_{\mu\nu}\partial^\mu h^{\alpha\nu}-\frac{c_3}{4}\partial_\alpha h \partial^\alpha h+\frac{c_5}{2}\partial_\mu h^\mu{}_\nu\partial^\nu h\,,
\ee
where $h=h^\mu{}_\mu$. This is nothing but the general quadratic action for a spin-2 field. In this theory there can be up to 10 propagating modes, but, as it is well-known, the theory must enjoy some gauge symmetries in order to avoid ghostly dofs. Before proceeding to that, let us notice that $c_2$ and $c_4$ appear degenerated, so that the linear order does not allow to completely fix the theory from consistency arguments. Furthermore, the normalisation of $h_{\mu\nu}$ allows to absorb one of the parameters (up to a sign). Irrespectively of the number of propagating modes, they all trivially propagate on the light cone owed to the Lorentz invariance of the background and the absence of any mass parameters. For Fourier modes of momentum $\vec{k}$ it is convenient to decompose $h_{\mu\nu}$ into helicity modes with respect to $\vec{k}$. Then, the helicity-1 sector will contain a ghostly mode unless the gauge symmetry $h_{\mu\nu}\rightarrow h_{\mu\nu}+2\partial_{(\mu} \xi_{\nu)}$ with $\partial_\mu \xi^\mu=0$ is imposed, what is called transverse diffeomorphisms or TDiffs. This symmetry restricts the parameters in (\ref{h_action}) to satisfy $c_2-c_1=2(c_4-1)$, which is of course fulfilled by the STEGR. In order to end up with two propagating dofs (as it corresponds to a massless spin 2 field) we need to complete the TDiffs to a four-parameter gauge symmetry, what can be achieved in two ways. The first possibility is to complete the symmetry to full linearised diffeomorphisms (Diffs$_{\rm CG}${\footnote{We denote this symmetry Diffs$_{\rm CG}$ as the Diffs symmetry that arises in the coincident gauge in order to distinguish it from the original Diffs that in turn are used to go to the coincident gauge.})  $h_{\mu\nu}\rightarrow h_{\mu\nu}+2\partial_{(\mu} \xi_{\nu)}$ with no constraints on $\xi^\mu$. This leads to additional constraints $c_5=c_3=c_1$, which indeed reproduce the values of the STEGR. The second possibility is to impose an additional Weyl symmetry (WTDiffs$_{\rm CG}$) $h_{\mu\nu}\rightarrow h_{\mu\nu}+\phi\eta_{\mu\nu}$ with $\phi$ an arbitrary scalar field. This symmetry further requires $c_3=3c_1/8$ and $c_5=c_1$. This is the linearised version of unimodular gravity, which differs from GR in the appearance of a cosmological constant as an integration constant. The general quadratic theory within the symmetric teleparallelism framework for this choice of parameters is yet to be analysed.

Let us remark that the above constraints are of paramount importance for the consistency of the theory so that, theories that fail to satisfy them will be prone to ghost-like instabilities. This is in fact a general result not only applicable to the quadratic theory but to a general non-linear extension\footnote{By non-linear extension we refer to the corresponding field equations not being linear in the non-metricity.} theory with Lagrangian $\Lag=f(Q_1,Q_2,Q_3,Q_4,Q_5)$ with $Q_i$ the five independent terms of the quadratic theory. Around a Minkowski background solution (provided such a solution exists), the quadratic Lagrangian for the perturbations will take the same form as (\ref{h_action}), with $c_i$ given in terms of $\partial f/\partial Q_i$. Thus, all these theories will be constrained by stability around Minkowski very much like the general quadratic theory. In particular, this crucially impacts the number of possible stable propagating polarisations in a general symmetric teleparallel theory.  Moreover, even if the linear perturbations succeed in fulfilling the stability conditions, the loss of gauge symmetries\footnote{Let us stress that all these theories are Diffs invariant by construction and, consequently, the coincident/unitary gauge also exists for them. The gauge symmetries we refer to here are the Diffs$_{\rm CG}$ that remain in the coincident gauge and that ensure the absence of additional propagating dofs.} when considering interactions will likely re-introduce ghostly degrees of freedom that were removed from the quadratic spectrum. This is in fact a very strong constraint that must be carefully taken into account for the theories to be consistent.

\subsection{$f(\QQGR)$ extensions}
A special case of a non-linear extension is given by $\Lag=f(\QQGR)$, which trivially fulfils the stability requirements around Minkowski because the only effect will be a re-scaling of the gravitational constant determined by $f'$. It is important to notice that, by virtue of (\ref{eq:RtoQ}), even if $\Lag_{\rm CGR}=\QQGR$ is equivalent to $\Lag_{\rm GR}=\mR$ because they only differ by a total derivative, precisely this boundary term makes $f(\mR)$ and $f(\QQGR)$ completely different. 

This specific non-linear extension, besides being a priori less prone to instabilities than the general non-linear extensions, exhibits  one of the crucial features of these extensions, namely, since Diffs$_{\rm CG}$ are only realised up to a total derivative in the STEGR, the $f(\QQGR)$ theories will no longer realise this symmetry\footnote{It can happen that a subset of Diffs$_{\rm CG}$ remain a symmetry. The conditions would be that such subset of symmetries give rise to a trivial boundary term. We will see an explicit example of this below.}, with important consequences. In order to illustrate some of these consequences and further motivate the special case represented by $f(\QQGR)$ among all the theories based on $f(Q_1,Q_2,Q_3,Q_4,Q_5)$ we will consider a cosmological background described by a FLRW metric with spatially flat sections
\be
\diff s^2=-N^2(t)\diff t^2+a^2(t)\diff\vec{x}^2
\ee
with $N(t)$ and $a(t)$ the lapse function and the scale factor, respectively. If we work in the coincident gauge, then we have exhausted all the freedom in choosing the coordinates so that, in principle, it is not legitimate to fix the lapse $N(t)$ to any particular value by means of a time reparameterisation, as it is usually done. However, the special case of $f(\QQGR)$ does permit to fix the lapse because the background action in the minisuperspace
\be
\mS=-\frac{1}{16\pi G}\int\diff^4x\sqrt{-g}f(\QQGR)=-\frac{1}{16\pi G}\int\diff^3x\diff t Na^3f\left(\frac{6\dot{a}^2}{a^2N^2}\right)
\ee
retains a time-reparameterisation invariance $t\to\zeta(t)$,  $N(t)\to N(t)/\dot{\zeta}(t)$ for an arbitrary function $\zeta(t)$. The gravitational field equations for $N=1$ and in the presence of a perfect fluid with density $\rho$ and pressure $p$ then are given by
\begin{align}
6f'H^2-\frac{1}{2}f &=  8\pi G\rho\,, \label{fried1} \\
\left( 12f''H^2 + f'\right)\dot{H} &=  -4\pi G\left(\rho+p\right)\,. \label{fried2}
\end{align}
A remarkable class of theories is given by $f=\QQGR+\Lambda\sqrt{\QQGR}$ with $\Lambda$ some parameter. This family is special because it gives exactly the same background evolution as GR irrespectively of $\Lambda$, which will thus only affect the evolution of the perturbations. Since the background equations of motion  (\ref{fried1}), (\ref{fried2}) are the same as those of the $f(\TTGR)$ theories, we will not go into more details here, but obviously the same cosmological solutions will be possible. The differences will arise in the perturbations. Going back to the existence of a time-reparameterisation symmetry in this specific theory, there will be  some associated Bianchi identities that applied to (\ref{fried1}), (\ref{fried2}) give
\be
\dot{\rho}+3H(\rho+p)=0\,,
\ee
completely consistent with the continuity equation of the matter sector. In order to show that this is a non-trivial result, we can consider the general quadratic theory, whose action in the minisuperspace of a FLRW universe is given by
\be
\mS=-\frac{1}{16\pi G}\int\diff^3x\diff t\frac{a^3}{N}\left[3\big(c_1-3c_3\big)\frac{\dot{a}^2}{a^2}+\big(c_1-2c_2-c_3+2c_4-2+2c_5\big)\frac{\dot{N}^2}{N^2}+6\big(c_5-c_3\big)\frac{\dot{a}\dot{N}}{aN}\right]
\ee
which clearly does not have the symmetry under time reparameterisations and the lapse is an additional dynamical degree of freedom. Thus, fixing the lapse is not legitimate and can lead to inconsistent equations of motion. More precisely, setting the lapse would be a selection of some particular branches of solutions, which are not guaranteed to exist a priory. This is related to the choice of good versus bad tetrads in the $f(\TTGR)$ case, where it was noted that some choices of tetrads led to inconsistent equations of motion, which is nothing but a reflection of overfixing a gauge.

The cosmological perturbations of the $f(\QQGR)$ theories will give crucial signatures for the discrimination of these theories. We will not go into the details of the perturbations equations, but will simply point out an interesting general feature that, in turn, may point towards the inviability of the whole family of theories. As we have repeatedly commented, we no longer have the freedom to choose the coordinates once we work in the coincident gauge. At the background level, time reparameterisations remains as a symmetry, but at the perturbative level there are no remanent gauge symmetries in general so we have to work with all the metric perturbations\footnote{Another possibility would of course be to re-introduce the connection through $\xi^\alpha$ and seek for a more convenient gauge involving both metric and connection perturbations.}. We will focus here on the scalar sector, so the metric will be decomposed as
\be
\diff s^2=-a(t)^2(1+2\phi)\diff \tau^2+2a^2\beta_{,i}\diff \tau \diff x^i+a^2\left[(1-2\psi)\delta_{ij}+2\left(\delta^k_i\delta^l_j-\frac12\delta^{kl}\delta_{ij}\right)\sigma_{kl}\right]\diff x^i \diff x^j
\ee
with $\phi$, $\psi$, $\beta$ and $\sigma$ the corresponding scalar potentials. As a remnant of the STEGR, the potentials $B$ and $\phi$ remain non-dynamical for the $f(\QQGR)$ theories and, therefore, they can be integrated out. We are then left with two dynamical scalar potentials. A very interesting feature of the perturbations that is worth mentioning here is their behaviour under a gauge transformation. Since this is no longer a symmetry, obviously, under a gauge transformation with parameters $\delta x^\mu=(\epsilon^0,\delta^{ij}\partial_j\epsilon)$ of the scalar potentials
\bea
\phi & \rightarrow & \phi -({\epsilon^0})' - \mathcal{H} \epsilon^0\,, \label{cgt1} \\
\beta & \rightarrow & \beta + \epsilon' + \epsilon^0\,, \\
\varphi & \rightarrow & \varphi - \frac{1}{3}\delta^{ij}\epsilon_{,ij} + \mathcal{H}\epsilon^0\,, \\
\sigma & \rightarrow & \sigma + \epsilon\,,
\eea
the equations will not be invariant. However, for the particular case of maximally symmetric backgrounds, i.e., Minkowski, de Sitter and anti de Sitter, there is a residual symmetry provided the gauge parameters satisfy $\xi^0+\xi'=0$. This means that these backgrounds will exhibit one less propagating mode, as can be also directly seen from the fact that the Hessian around these backgrounds becomes degenerate. This feature might however signal the potential presence of a strong coupling problem for these backgrounds, since this symmetry would seem accidental and, in any case, these backgrounds would seem to present a discontinuity in the number of propagating dofs. This strong coupling problem may represent a fatal flaw of these theories since Minkowski and/or de Sitter are desirable stable background solutions.

\section{Matter couplings}
\label{matter}

Besides the purely gravitational sector, prescribing how matter couples is a foundational aspect of gravity. The majority of the matter fields living on a manifold with a general connection will be oblivious to the presence of the distortion. However, in order to rigorously investigate the possible existence of subtleties, one has to be aware of the fact that
\begin{itemize}
\item generalized geometries give room for ambiguity in the matter coupling;
\item crucial differences arise for bosonic and fermionic fields.
\end{itemize}
In consideration of the first point, let us remind ourselves that if one considers the action of a point particle $\mathcal{S}=-mc^2\int d\tau$, this particle will only access the Levi-Civita part of the connection. In GR this is equivalent to postulating that the point particle will follow the geodesic equation $\frac{d^2x^\mu}{d\tau^2}+\Gamma^\mu{}_{\nu\alpha} \frac{dx^\nu}{d\tau}\frac{dx^\alpha}{d\tau}=0$ with $\Gamma^\mu{}_{\nu\alpha}=\{^\mu_{\nu\alpha}\}$. However, in generalized geometries starting from the action or the postulated geodesic equation will not give rise to the same conclusion and introduce ambiguities.

Concerning the second point, crucial differences arise for bosonic and fermionic fields because bosons only couple to the metric, but fermions also couple to the connection. Already within GR, it is necessary to introduce an additional structure in order to define spinors in curved spacetimes: the tetrads. Another crucial point is whether one starts from the minimal coupling procedure. Bosonic particles minimally coupled to gravity, with the prescription $\eta_{\mu\nu} \rightarrow g_{\mu\nu}$, $\diff \rightarrow \diff$, will only see the Levi-Civita part of the connection. Hence, they will follow the above geodesic equation with $\Gamma^\mu_{\nu\alpha}=\{^\mu_{\nu\alpha}\}$. Starting from the geodesic equation, it is clear that the torsion does not contribute since the geodesic equation is symmetric under the exchange of $\nu \leftrightarrow \alpha$. However, the minimal coupling prescription $\eta_{\mu\nu} \rightarrow g_{\mu\nu}$, $\diff \rightarrow \Diff$, the latter implying that $\partial_\mu \rightarrow \nabla_\mu$, can already be problematical for bosonic fields of nonzero spin. For example, the gauge invariance of the Maxwell field $A^\mu$ would need to be reconsidered in TEGR due to the appearance of a non-gauge invariant coupling to torsion in $F_{\mu\nu}=2\nabla_{[\mu}A_{\nu]}$. 

In general, fermions will be very sensitive to the presence of any distortion of the connection. The TEGR encounters some difficulties in coupling gravity to fermions because the natural coupling is to the Weitzenb\"ock connection \cite{Obukhov:2004hv}.

The STGR elegantly avoids this difficulty due to the absence of torsion and the fact that the Dirac Lagrangian is blind to the non-metricity so that fermions are only concerned with the usual Levi-Civita piece of the connection \cite{BeltranJimenez:2018vdo}.

In all these formulations, the corresponding dynamics of the matter fields will non-trivially depend on the assumed matter action and whether the minimal coupling prescription is selected on purpose. This is also the case in the standard formulation of GR and the choice has to be done based on the wanted physical effects.
\subsection{General Relativity}\label{subsecMatterGR}
When we write down the Hilbert action in the presence of the standard matter fields
\be
\mathcal{S}_{\rm GR_{(2)}}=\frac{1}{16\pi G}\int\diff^4x\sqrt{-g}\,\mathcal{R}(g)+\mathcal{S}_{\rm matter}[g_{\mu\nu},\phi]\,,
\label{eq:EHactionMatter}
\ee
there enter already non-trivial assumptions on the physical system at hand. Namely, one has explicitly assumed that
\begin{itemize}
\item the matter fields do not couple do the connection, and
\item the minimal coupling prescription $\eta_{\mu\nu} \rightarrow g_{\mu\nu}$, $\partial_\mu \rightarrow \nabla_\mu$ with $\Gamma^\mu{}_{\nu\alpha}=\{^\mu_{\nu\alpha}\}$ is applied.
\end{itemize}
Therefore, only the variation with respect to the metric has to be performed, yielding 
\be
G_{\mu\nu}=\mathcal{R}_{\mu\nu}-\frac12\mathcal{R}g_{\mu\nu}=\frac{T_{\mu\nu}}{\mpl^2},
\ee
with the stress energy tensor defined as
\begin{equation}
T_{\mu\nu}=\frac{-2}{\sqrt{-g}}\frac{\delta \mathcal{S}_{\rm matter}}{\delta g^{\mu\nu}}\,.
\end{equation}
The Bianchi identities, i.e. the divergenceless nature of the Einstein tensor, enforces $\nabla^\mu T_{\mu\nu}=0$ upon the matter fields. This is directly related with the consistency of the matter fields equations of motion. If one instead assumes the starting Lagrangian to be of the form 
\be
\mathcal{S}_{\rm GR_{(1)}}=\int\diff^4x\left[ \frac{\sqrt{-g}}{16\pi G}R(g,\Gamma) + \lambda_{\alpha}{}^{\mu\nu}T^\alpha{}_{\mu\nu} + \hat{\lambda}^{\alpha}{}_{\mu\nu}Q_{\alpha}{}^{\mu\nu}\right]+\mathcal{S}_{\rm matter}[g_{\mu\nu},\Gamma^\alpha{}_{\mu\nu},\phi]\,,
\label{eq:EHaction1matter}
\ee
with an explicit coupling of the matter field to the general connection, then the variation of the action with respect to the connection yields
\begin{eqnarray}\label{EOM_connection}
\nabla_\rho\left[\sqrt{-g}g^{\mu\nu}\right]-\delta^\mu_\rho\nabla_\alpha\left[\sqrt{-g}g^{\alpha\nu}\right]=\sqrt{-g}\left[ g^{\mu\nu}T^\alpha_{\alpha\rho}+g^{\alpha\nu}T^\mu_{\rho\alpha}-\delta^\mu_\rho g^{\beta\nu}T^\alpha_{\alpha\beta} \right]+\Delta^{\mu\nu}_\rho\,,
\end{eqnarray}
with the hypermomentum of the matter fields defined as
\be
\Delta^{\mu\nu}_\rho=\frac{2}{\mpl^2}\frac{\delta \mathcal{S}_{\rm matter} }{\delta\Gamma^\rho{}_{\mu\nu}}\,,
\ee
which arises due to the coupling to the connection. Given the torsion-free and metric constraints enforced by the Lagrange multipliers, Eq. (\ref{EOM_connection}) would imply that the hypermomentum must be identically zero, giving rise to non-trivial constraints for the matter fields, specially fermions, which do carry hypermomentum due to their coupling to the axial torsion. Thus, when including matter fields, we must either consider minimally coupled fields or formulate the theory in an unconstrained metric-affine formalism for the consistency of the theory.

Already in the standard formulation of GR the presence of fermions requires the introduction of a vielbein and the gravitational spin connection. The information about the spacetime and the spin meets in the Clifford algebra with the Dirac matrices acting at each spacetime point. Dirac's equation in curved spacetime then naturally takes the form
 \be
 i\gamma^a \ie^\mu_a D_\mu \Psi-m\Psi=0
 \ee
 with the Dirac matrices $\gamma^a$ and the covariant derivative $D_\mu=\partial_\mu-\frac{i}{4}w_\mu^{ab}\sigma_{ab}$ defined in terms of the spin connection $w_\mu^{ab}$ and 
 $\sigma_{ab}=\frac{i}2[\gamma_a,\gamma_b]$. This equation follows naturally from the minimal coupling prescription discussed above\footnote{It is worth stressing that the equation obtained from the covariantization of the Dirac Lagrangian does not give the covariantized version of the Dirac equation in spaces with torsion and/or non-metricity. This subtlety is irrelevant in the case of GR, but it is important when considering more generally connected spacetimes.}.
In this way, the vielbein approach supports a local symmetry of Lorentz transformations in tangential space and diffeomorphism invariance. Hence, already within GR, fermions need a special care and the introduction of an additional structure.

\subsection{Metric teleparallelism}

It is possible to stipulate that for the bosonic fields, the minimal coupling is $\partial_\mu\rightarrow \partial_\mu$, whilst for the fermionic fields one sets $\partial_\mu\rightarrow \nabla_\mu$.
This may seem arbitrary in view of that gravitation is a universal force, under which also bosonic fields are ``charged'' in principle. However, adopting the covariant prescription 
for bosons as well as fermions will lead to problems with gauge fields in metric teleparallelism. To see this, consider the simplest example, the photon $A_\mu$, whose field strength in the absence of gravitation is $F_{\mu\nu}=2\partial_{[\mu} A_{\nu]}$, and would become $F_{\mu\nu} \rightarrow 2\nabla_{[\mu} A_{\nu]} = 2\partial_{[\mu} A_{\nu]} + T^\alpha{}_{\mu\nu}A_\alpha$ in the universal covariant prescription. In the case of both GR and STGR nothing happens to the Maxwell field strength, since there is no torsion. However, in the context of metric teleparallelism this is obviously not the case, and the photon becomes non-minimally coupled to torsion. That spoils the $U(1)$ invariance, at least in its standard form.

The problem of coupling fermions in metric teleparallelism is seen easily from the definition of the spin connection given above, $D_\mu=\partial_\mu-\frac{i}{4}w_\mu^{ab}\sigma_{ab}$. When 
$w_\mu^{ab}$ is the pure-gauge connection that can be always be set to zero by a Lorentz rotation, the fermions are obviously decoupled from the Levi-Civita connection (\ref{levi-civita}). However, that coupling would be required to ensure the usual energy-momentum conservation $\mathcal{D}_\mu T^\mu{}_\nu$. The problem can of course be avoided by re-invoking the GR coupling  $\partial_\mu \rightarrow \mathcal{D}_\mu$ now in metric teleparallelism. One at least heuristic justification for such a prescription is that by writing the pure-gauge $w_\mu^{ab}$ in terms of trivial tetrads and then promoting those to the full tetrads indeed would make the pure-gauge spin connection become the metric spin connection of GR \cite{Aldrovandi:2013wha,Krssak:2018ywd}. However, this is not the standard procedure in gauge theories. Thus, the conventional coupling principle $\partial_\mu\rightarrow \nabla_\mu$ in metric teleparallelism is not viable for either bosons or fermions. 

\subsection{Symmetric Teleparallelisms}
The coupling to matter within the realm of symmetric teleparallel theories can be performed following the usual minimal coupling prescription
\be \label{STGRmatter}
\mS_{\rm STEGR}=-\int\diff^4x\left[\frac{1}{16\pi G}\sqrt{-g}\,\QQGR+\lambda_{\alpha}{}^{\beta\mu\nu} R^\alpha{}_{\beta\mu\nu}+\lambda_\alpha{}^{\mu\nu} T^\alpha{}_{\mu\nu}\right]+\mathcal{S}_{\rm matter}[g_{\mu\nu},\phi,\nabla_\mu\phi]\,.
\ee
This theory is then equivalent to GR where the matter fields in $\mathcal{S}_{\rm matter}$ follow the same physical geodesic equations as in GR, since the couplings are exactly the same for the known fields in the Standard Model. For bosonic fields, we will then have $\nabla_\mu\phi\rightarrow\partial_\mu\phi$ as in the usual formulation of GR. There is a subtle point concerning the general quadratic theory that is worth explaining in detail. In that case, the connection field equations do not trivialise in the coincident gauge (as they do for the Coincident GR). However, the corresponding equations can be obtained by applying the Bianchi identities to the metric field equations and, thus, they are redundant with them. Hence, the information in the connection field equations is of course not lost when working in the coincident gauge in general theories.

In difference to the previous reformulation, in STGR fermions do not require any adjustments to the minimal prescription. The standard derivative $\partial_\alpha$ of the usual flatspace Dirac Lagrangian is the same as $\nabla_\alpha$ (in the coincident gauge we simply have $\nabla_\alpha=\partial_\alpha)$. In order to appreciate this statement, bear in mind that even 
if the covariant derivative $\nabla_\alpha$ appears in the action, only $\mathcal{D}_\alpha$ survives 
in the equations of motion for the Dirac field due to the Hermitean property of the 
Dirac action. Let us remind ourselves that $\Gamma^\alpha{}_{\mu\nu} =
\left\{^{\phantom{i} \alpha}_{\mu\nu}\right\} + L^\alpha{}_{\mu\nu} = 0$ in the coincident gauge, from which the piece $L^\alpha{}_{\mu\nu}$ is filtered out due to the 
Hermitean nature of the action\footnote{To see how this occurs, first generalise the Lorentz basis $\sigma_{ab}=\frac{i}2[\gamma_a,\gamma_{b}]$ to the general linear basis, 
$\frac{i}2\gamma_a\gamma_b=\sigma_{ab} - i\eta_{ab}$. Thus only the trace $Q_\mu$ of non-metricity contributes to the covariant derivative of spinors in the first place.
Secondly, in the Dirac action this trace simply cancels due to the Hermiteanisation. As the result, even though in the coincident gauge we have $\partial_\mu\rightarrow \nabla_\mu=\partial_\mu$ in the action, effectively we recover $\partial_\mu \rightarrow \mathcal{D}_\mu$ in the equations of motion.} .

Therefore, the standard equation of motion in curved spacetime
arises for the spinors, which only involves the $\left\{^{\phantom{i} \alpha}_{\mu\nu}\right\}$ part. Hence, the Dirac fields are completely oblivious to any disformation of geometry given by a general  $Q_{\alpha\mu\nu}$. 

Needless to say that if we consider couplings beyond the minimal coupling procedure where for instance matter fields could directly explicitly couple to the connection, there will be notable physical effects beyond GR due to the presence of the hypermomentum of the matter fields.

\section{Conclusions}

\begin{figure*}[htbp]
\centering
\includegraphics[width=16cm]{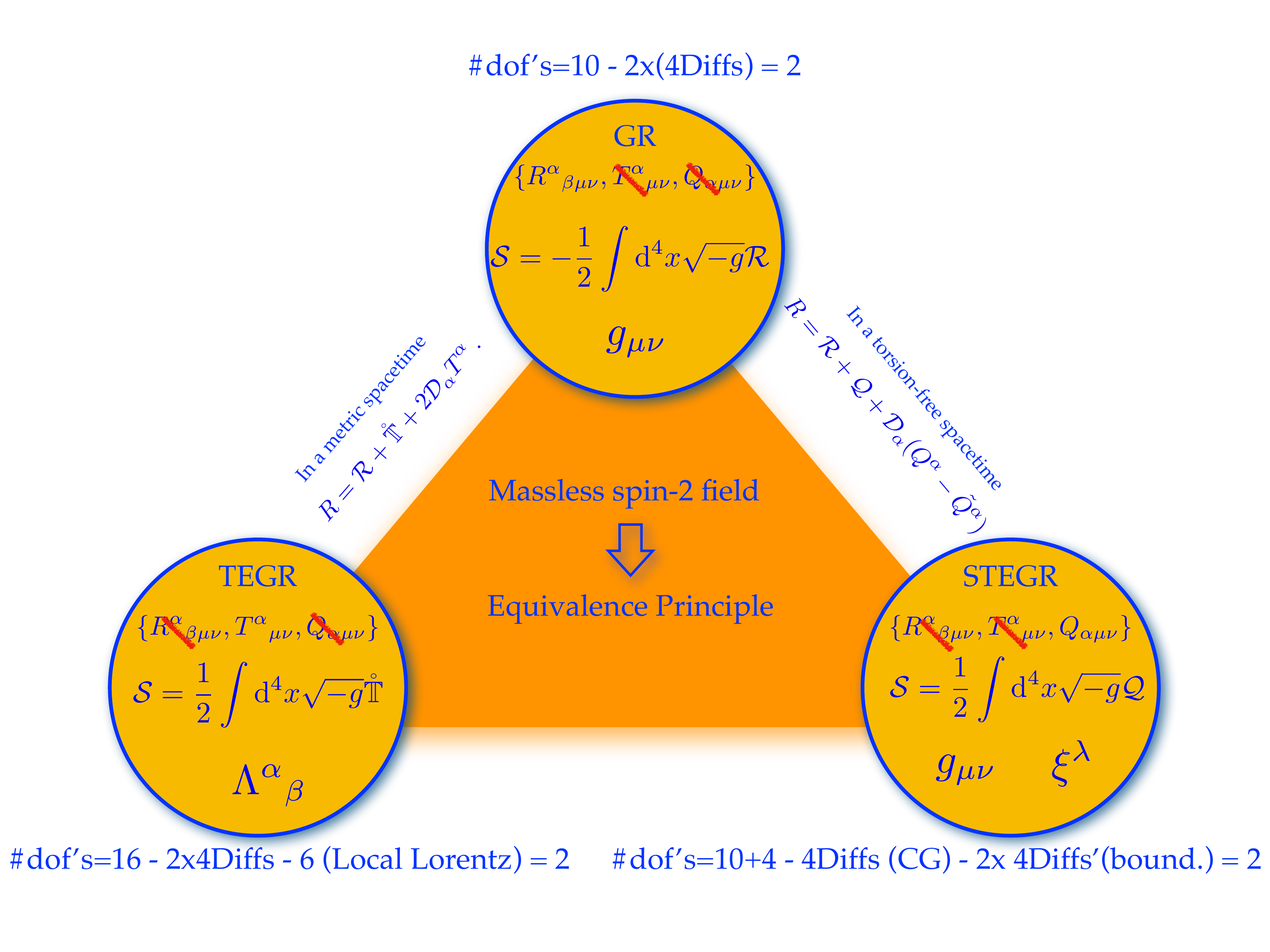}
\caption{This figure summarises the three alternative gravitational descriptions of gravity together with its main properties. In the GR description, the fundamental object is the metric tensor $g_{\mu\nu}$, the spacetime is curved, but metric-compatible and torsion-free and the 4 Diffs symmetry reduce the dofs from 10 to 2. The TEGR has the inertial connection parameterised by $\Lambda^\alpha{}_\beta\in GL(4,\mathbb{R})$ as its fundamental object, which generates torsion, but the connection and the non-metricity are trivial. Diffs plus the local Lorentz symmetry reduce the 16 independent components of $\Lambda^\alpha{}_\beta$ to 2. Finally, the STEGR contains the metric $g_{\mu \nu}$ and $\xi^\alpha$ as fundamental elements. In this case, $\xi^\alpha$ generates a flat and torsion-free connection and only the non-metricity piece is left. Furthermore, $\xi^\alpha$ can be fully removed by a suitable choice of coordinates (the coincident gauge) leaving a trivial connection. In that gauge, the presence of a second Diffs symmetry realised up to boundary terms reduce the number of dofs to two. At the heart of these equivalences lyes the fact that gravity describes a theory for an interacting massless spin-2 particle, whose consistency requires the equivalence principle and, thus, the possibility of describing it in geometrical terms.}
\label{FigA}
\end{figure*}

The ternion of geometrical representations of GR offers useful complementary perspectives to the theory of gravity. The non-trivial boundary terms that differentiate the three formulations present a new tool to explore the holographic nature of GR. In these notes, we have reviewed the formulation of GR in three classes of geometries, and wrote down the six actions indicated below.
\begin{itemize}
\item  Riemannian: in terms of the general (\ref{eq:EHaction1}) and the metric  (\ref{eq:EHaction}) connection.
\item  Metric teleparallel: in terms of the general (\ref{abcd}) and the inertial (\ref{ab}) connection.
\item  Symmetric teleparallel: in terms of the general (\ref{abcd2}) and the inertial (\ref{Eq:CGRaction}) connection.
\end{itemize}
The geometrical trinity that emerges is depicted in Figure \ref{FigA}. We also considered briefly the perspective from the frame bundle in Section \ref{vierbein}, and discussed some of the most straightforward generalisations of the two versions of teleparallel GR.

Symmetry is foundational to theoretical physics, but the geometry chosen for its illustration may be a matter of convention. 
From the perspective of gauge theory, we understand that spin connection
is the gauge potential of Lorentz rotation and the curvature is its gauge field strength, whilst the tetrad is related to the gauge potential of translation and the torsion to its gauge field strength. Gravitation can be geometrised in terms of either of these, and in fact Einstein considered both of the corresponding mathematical formulations in time. However, to him the main achievement of GR was never the geometrization of gravity {\it per se} but its unification with inertia\footnote{See Ref. \cite{history} for an elaboration on this point and references.}. It is the essence of this unification, as expressed in the equivalence principle, that gravitation can always be locally eliminated by changing the coordinate system. At the same time, it is a basic fact about gauge theories that a gauge field force can be made to locally vanish if it has zero field strength. We may speculate\footnote{Recall indeed from (\ref{el}) that the theory is the minimal covariantization of the Einstein Lagrangian (1916).} that the coincident GR, which purifies gravity from both torsion and curvature, would have been the ``Einstein's third GR'' had he lived long enough to witness the spectacular success of the gauge principle in the theories of particle physics. 

The coincident GR realises gravity as the gauge theory of the group of translations, which is the natural interpretation for the universal interaction sourced by energy and momentum, the conjugates of the time and space translations, respectively.  
The metric teleparallel torsion theory also had been suggested as a gauge theory of the translation group \cite{Aldrovandi:2013wha}. From the gravity side, this interpretation however fails in that the
connection is not a translation but a Lorentz rotation as clarified\footnote{More precisely, the spin connection $\omega$ is an $SO$, the affine connection $\Gamma$ is a $GL$ transformation.} in Section \ref{tele}; from the matter side, the interpretation fails due to the inconsistency of the minimal coupling that
was discussed in Section \ref{matter}. A paradox about the Diffs is that the consistent realisation of their underlying gauge theory does not allow the corresponding gauge field strength to exist.   
This reflects the special property of the gravitational interaction whose ``external'' gauge geometry describes the spacetime itself, the arena for the compact, ``internal'' geometries that describe the
interactions of matter fields in the standard model of particle physics.

Nevertheless, all of the three representations of the geometrical trinity remain to be useful and provide important complementary perspectives to the nature of gravity.

\acknowledgments 
We are grateful to the $3^{rd}$ Jos\'e Pl\'inio Baptista School on Cosmology held in Pedra Azul and their participants for a magnificent atmosphere. JBJ acknowledges support from the  {\it Atracci\'on del Talento Cient\'ifico en Salamanca} programme and the MINECO's projects FIS2014-52837-P and FIS2016-78859-P (AEI/FEDER). LH is supported by funding from the European Research Council (ERC) under the European Unions Horizon 2020 research and innovation programme grant agreement No 801781 and by the Swiss National Science Foundation grant 179740.


\end{document}